\documentclass[1p]{elsarticle}

\usepackage[T1]{fontenc}
\usepackage[utf8]{inputenc}
\usepackage{geometry}
\usepackage{amsmath,amssymb}
\usepackage{graphicx}
\usepackage{booktabs}
\usepackage{algorithm}
\usepackage{algpseudocode}
\usepackage{url}
\usepackage[colorlinks=true,linkcolor=blue,citecolor=blue,urlcolor=blue]{hyperref}

\begin{document}

\begin{frontmatter}

\title{Co-occurrence Patterns of LoRA Adapters in Production Diffusion Model Inference Services}

\author[1]{Tao Zhang}
\author[2,3]{Bin Liao}\cortext[cor1]{Corresponding author.}
\author[4]{Tao Zhou}
\author[2,3]{Yanping Liu}

\affiliation[1]{organization={School of Information Engineering, Guizhou University of Traditional Chinese Medicine}, city={Guiyang}, postcode={550025}, country={China}}
\affiliation[2]{organization={School of Big Data and Statistics, Guizhou University of Finance and Economics}, city={Guiyang}, postcode={550025}, country={China}}
\affiliation[3]{organization={School of Mathematics and Statistics, Guizhou University of Finance and Economics}, city={Guiyang}, postcode={550025}, country={China}}
\affiliation[4]{organization={School of Mathematics and Statistics, Beijing Technology and Business University}, city={Beijing}, postcode={100048}, country={China}}

\begin{abstract}
Low-rank adaptation (LoRA) has become a key technology for serving large-scale personalized large language models and diffusion models in the cloud. However, the co-occurrence patterns, resource contention relationships, and evolutionary regularities of adapters under production inference workloads have not been systematically or quantitatively studied. Based on GenTD26, Alibaba's production diffusion model inference dataset, this paper adopts a graph-theoretic framework to construct an adapter co-occurrence network and conducts a characterization from both static structure and dynamic evolution. Our main findings are as follows. (1) The co-occurrence network is extremely sparse, and adapter usage frequency follows a significant heavy-tailed distribution. (2) Introducing the first adapter incurs a 66.1\% execution-latency overhead, with diminishing marginal costs afterwards. (3) Co-occurrence relationships are driven by base models: in 90.6\% of multi-adapter requests, all adapters share the same dominant base model; 66.2\% of significant co-occurrence edges connect same-model adapter pairs; and in 85.8\% of multi-adapter requests, all adapter pairs form significant co-occurrence edges. (4) The adapter ecosystem exhibits a core--periphery bipolar structure, with a weekly Jaccard similarity of 0.696 at the model level and a churn rate of 54.5\% for the top-10 hottest models within a 12-hour window. Based on these findings, we propose a preloading strategy built on top-$k$ co-occurrence statistics; offline experiments show that it covers 81.0\% of test-set co-occurrence pairs at $k=3$, and sensitivity analyses across frequency thresholds and time windows verify the robustness of the conclusions. These results provide a data-driven basis for cache preloading, adaptive scheduling, and GPU memory management in LoRA inference services.
\end{abstract}

\begin{keyword}
LoRA adapter \sep co-occurrence network \sep diffusion model inference \sep workload characterization \sep GPU cluster trace
\end{keyword}

\end{frontmatter}

\section{Introduction}
\label{sec:intro}

Diffusion models, represented by Stable Diffusion, have become the core engine of commercial image generation services, processing millions of user requests per day on large GPU clusters \cite{ref1}, and have found wide application in fields such as medical data generation \cite{ref3} and personalized image generation \cite{ref4}. Low-Rank Adaptation (LoRA) \cite{ref2} attaches lightweight adapter weights so that a single base model can support thousands of personalized variants, significantly reducing storage and computation costs while greatly improving the deployment flexibility of customized services. However, this ``one-base-multiple-adapters'' architecture also poses structural challenges to GPU resource management: inference requests must dynamically load one or more adapters on top of the base model, and the loading latency, GPU memory footprint, and scheduling order of adapters are mutually coupled, making request-level resource demands highly time-varying \cite{ref5,ref6}.

Unlike traditional inference services that run fixed model architectures \cite{ref12,ref13}, the LoRA adapter ecosystem exhibits a distinct long-tail usage pattern. The vast majority of adapters serve niche customization scenarios; only a very small number appear frequently, and multiple adapters may be invoked simultaneously by the same request. This workload characteristic raises a series of key system design questions: How do adapters co-occur within requests? Are co-occurrence relationships random, or do they exhibit predictable structural regularities? Which adapters should be preferentially resident in the GPU cache, and which can be loaded on demand? And are co-occurrence patterns stable over time? The answers to these questions directly determine the design of cache preloading, GPU memory allocation, and request batching and scheduling.

Existing work on LoRA inference systems mainly focuses on architecture and mechanism design. Punica \cite{ref7} and S-LoRA \cite{ref8} design sharded memory management and on-demand adapter loading mechanisms, significantly increasing the number of concurrent adapters that a single GPU can host; MixLoRA \cite{ref9} optimizes GPU utilization for concurrent execution of multiple adapters; and CaraServe \cite{ref10} explores CPU-assisted rank-aware loading to reduce adapter-switching overhead. When designing scheduling policies, these systems typically assume that adapter access patterns are known or adopt general-purpose caching strategies such as LRU, lacking structural characteristics of real workloads as a design basis. The Rock system \cite{ref6} conducts adapter popularity and burstiness analysis based on the GenTD26 trace, but its characterization remains at the aggregate statistical level and does not reveal the co-occurrence topology and structural properties among adapters. The SoCC'25 paper \cite{ref5} provides a top-down characterization of diffusion model inference services but does not analyze the co-occurrence structure.

To address the above problems, this paper starts from the structural associations among adapters and studies the co-occurrence patterns of LoRA adapters in production diffusion model inference services through a graph-theoretic approach. Unlike existing work that focuses on aggregate statistics (such as occurrence frequency and request share), this paper focuses on the co-occurrence topology and structural evolution patterns among adapters. The main contributions of this paper are as follows.

(1) Co-occurrence network analysis. Based on production traces, we construct a LoRA adapter co-occurrence network and reveal its extreme sparsity (under the threshold $\tau_{\min}=10$, $|V|=274$, $|E|=68$, and an edge density of only 0.18\%) and its significant heavy-tailed frequency distribution. Through connected-component and community-detection analysis, we further identify the structural characteristic of ``a small number of clusters plus a large number of isolated nodes.''

(2) Multi-dimensional quantitative workload analysis. In addition to basic network structure analysis, this paper quantitatively analyzes the workload from four perspectives: Jaccard similarity, base model--adapter association, dynamic changes at three time scales (request level, model level, and cross-model level), and latency degradation.

(3) System design strategy analysis. The co-occurrence-based preloading strategy covers 81.0\% of test-set co-occurrence pairs at $k=3$ (with a range of [78.1\%, 81.3\%] over 5 random seeds), and in 85.8\% of multi-adapter requests all adapter pairs form significant co-occurrence edges, providing a data-driven basis for cache preloading; frequency--centrality correlation analysis can guide the setting of scheduling priorities under memory constraints.

The remainder of this paper is organized as follows. Section~\ref{sec:related} reviews related work; Section~\ref{sec:formal} formalizes the relevant problems; Section~\ref{sec:dataset} describes the dataset and presents its overall characteristics; Section~\ref{sec:network} analyzes the co-occurrence network structure; Section~\ref{sec:temporal} characterizes the dynamic evolution; Section~\ref{sec:design} discusses implications for system design; and Section~\ref{sec:summary} discusses limitations and concludes the paper.

\section{Related Work}
\label{sec:related}

This section analyzes research in two areas: LoRA adapter serving systems and the characterization of production workloads, and clarifies the differences between this paper and existing work.

\subsection{LoRA Adapter Serving Systems}

After the proposal of LoRA, efficiently serving a large number of adapters quickly became a research hotspot in the systems community. Punica \cite{ref7} implements efficient batching of different LoRA requests on a single GPU through segmented gather matrix-vector multiplication kernels; S-LoRA \cite{ref8} introduces a unified paging mechanism that manages the KV cache and LoRA weights together, supporting thousands of concurrent adapters; MixLoRA \cite{ref9} interleaves the computation of heterogeneous adapters within the same batch, improving GPU utilization. Reference~\cite{ref17} applies LoRA instruction fine-tuning to domain-specific training of multimodal large models, and Petals \cite{ref18} explores a distributed inference architecture for cross-node collaboration. In the broader area of generative AI inference optimization, systems such as vLLM \cite{ref19}, Splitwise \cite{ref20}, ServerlessLLM \cite{ref21}, and FlexPipe \cite{ref22} also continuously optimize inference efficiency from perspectives including GPU memory management, phase splitting, cold-start loading, and pipeline refactoring.

Regarding adapter loading overhead, recent work has begun to explore loading--computation orchestration and cache management. CaraServe \cite{ref10} adopts a CPU-assisted strategy that executes adapter computation in the prefill stage in parallel with GPU-side adapter loading, reducing cold-start latency; Toppings \cite{ref33} proposes a CPU-assisted, rank-aware scheduling algorithm that reduces average request latency by up to 1.7$\times$; AuLoRA \cite{ref34} orchestrates loading and computation at the layer granularity, improving GPU utilization through layer-priority loading and intra-layer pipeline execution. In cache management, Chameleon \cite{ref35} utilizes idle GPU memory to cache popular adapters and designs an adapter-aware scheduling strategy to minimize loading overhead; JointSerLoRA \cite{ref36} jointly considers adapter loading states and KV cache memory contention and designs a cache-aware scheduling mechanism. Notably, when designing cache and scheduling policies, the above systems typically assume that adapter access patterns are known or adopt general-purpose strategies such as LRU. They answer the question of ``how to execute efficiently after adapters are loaded,'' but not the prior question of ``which adapters will be requested together.''

\subsection{Characterization of Production Workloads}

Characterizing production cluster workloads is a foundational methodology in system resource management research. Since the analysis of early Google cluster traces \cite{ref11}, researchers have recognized that detailed workload measurement can reveal failure modes of general-purpose strategies in real environments; Resource Central \cite{ref12} further demonstrates that workload understanding can be directly translated into quantifiable scheduling benefits. This tradition has continued in GPU data center research: MLaaS \cite{ref13} characterizes an Alibaba production cluster with more than 6,000 GPUs; the Philly trace \cite{ref14} focuses on task arrival and GPU utilization characteristics of training workloads; Reference~\cite{ref15} systematically characterizes the resource demands and predictability of deep learning workloads; and Reference~\cite{ref16} provides a survey of scheduling research in GPU data centers. In addition, abstracting workloads into graph structures for analysis has been proven effective in revealing hidden dependency relationships in scenarios such as data centers (VM placement \cite{ref23}, task dependencies \cite{ref24}, microservice invocation \cite{ref25}) and machine learning workloads (model dependencies \cite{ref26}, interference patterns \cite{ref27}).

In the field of diffusion model inference services, the GenTD26 dataset \cite{ref5} records multi-component pipelines, LoRA adapter configurations, and full-stack performance data. Based on this dataset, the SoCC'25 paper \cite{ref5} provides a top-down characterization of inference services; the Rock system \cite{ref6} further characterizes adapter popularity and request burstiness, and achieves an 84.1\% cache hit rate based on heterogeneous-aware resource orchestration. SwiftDiffusion \cite{ref37} analyzes the inference request traces of a commercial text-to-image service and finds that add-on modules such as ControlNet and LoRA are widely used in production, and their loading overhead significantly affects service latency and GPU resource efficiency. The above work analyzes ``which adapters are most popular'' and ``how occurrence frequency is distributed,'' but does not analyze the co-occurrence relationships and structural regularities among adapters, which directly affect the design of cache preloading and request batching strategies.

Unlike the above work, this paper focuses on the co-occurrence topology and structural evolution patterns among adapters. Characterizing co-occurrence patterns through graph-theoretic methods answers the questions of ``which adapters appear together, in what structure they cluster, and whether they are stable over time.'' Specifically, in terms of analysis granularity, our analysis shifts from frequency statistics of individual adapters to the co-occurrence relationships among adapters, introducing co-occurrence network construction, connected components, and community detection. Co-occurrence regularities can be used to guide cache preloading and scheduling priority setting, and can provide references for workload optimization of other componentized services (such as plugin markets and dependency combinations in function-as-a-service).

\section{Problem Formalization and Co-occurrence Analysis Methods}
\label{sec:formal}

This section formalizes the graph-theoretic framework used to analyze LoRA adapter co-occurrence patterns. Let $A = \{a_1, a_2, \dots, a_N\}$ be the set of $N$ unique LoRA adapters observed in the production trace, and let $R = \{r_1, r_2, \dots, r_M\}$ be the set of $M$ requests. Each request $r_k$ is represented by a tuple $r_k = (t_k, b_k, A_k, l_k)$, containing the timestamp $t_k$, the base model $b_k$, the adapter set $A_k \subseteq A$, and the execution latency $l_k$ in seconds.

\subsection{Problem Formalization}

\textbf{Definition 1 (Co-occurrence event).} If a request simultaneously loads two different adapters, the two adapters are said to co-occur in that request. A request may trigger multiple co-occurrence events simultaneously.

\textbf{Definition 2 (Co-occurrence count matrix).} The co-occurrence matrix $C \in \mathbb{N}^{N \times N}$ has elements defined as the number of co-occurrences of any two adapters across all requests, as shown in Eq.~(\ref{eq:cooc}):

\begin{equation}
C_{ij} = \left| \{ r_k \in R : a_i \in A_k \land a_j \in A_k \land i \neq j \} \right|
\label{eq:cooc}
\end{equation}

The diagonal entry $C_{ii}$ represents the total number of occurrences of adapter $a_i$ (i.e., the number of requests containing $a_i$). To filter out incidental co-occurrences and focus on statistically meaningful relationships, we impose an edge weight threshold: edges satisfying $C_{ij} \geq w_{\min}$ (default $w_{\min} = 3$) are retained. We impose a frequency threshold $\tau_{\min}$ on nodes (default set to 10) and retain only adapters satisfying $C_{ii} \geq \tau_{\min}$, yielding the node set $V_{\tau} = \{ a_i \in A : C_{ii} \geq \tau_{\min} \}$. The co-occurrence network is thus defined as $G_{\tau} = (V_{\tau}, E_{\tau})$, where $E_{\tau} = \{ (i,j) : C_{ij} \geq w_{\min} \}$ and the edge weight is $w_{ij} = C_{ij}$. Throughout this paper, the default values are $\tau_{\min} = 10$ and $w_{\min} = 3$, and we perform sensitivity analysis over $\tau_{\min} \in \{5, 10, 20, 50\}$.

\subsection{Structural Metrics of the Co-occurrence Network}

(1) \emph{Network edge density}: defined as the ratio of the actual number of edges to the maximum possible number of edges. It characterizes the overall sparsity of the network and is the primary indicator for judging whether co-occurrence is a prevalent phenomenon. A value close to 0 indicates ``almost no co-occurrence,'' while a value close to 1 indicates ``co-occurrence everywhere.''

(2) \emph{Connected component}: defined as the maximal subset of nodes in which at least one path exists between any two nodes, i.e., a group of adapters that are mutually reachable through paths in the network. Connected components can reveal natural clusters among adapters.

(3) \emph{Jaccard similarity}: the number of co-occurrences of two adapters divided by the union of their respective occurrence counts. If two adapters always appear in pairs, the Jaccard value approaches 1; if they are each popular but rarely used together, the Jaccard value is small.

\begin{equation}
J_{ij} = \frac{|R_i \cap R_j|}{|R_i \cup R_j|} = \frac{C_{ij}}{C_{ii} + C_{jj} - C_{ij}}
\label{eq:jaccard}
\end{equation}

where $R_i = \{ r_k \in R : a_i \in A_k \}$. The closer $J_{ij}$ is to 1, the stronger the co-occurrence dependency. A permutation test (e.g., 1,000 random shuffles of adapter labels) is used to determine whether the observed mean significantly deviates from the random expectation.

\subsection{Co-occurrence Strength and Structural Characteristics}

(1) \emph{Modularity $Q$ of partition $C$}: as a quantitative verification of the degree of fragmentation, we adopt greedy modularity maximization~\cite{ref28} to detect community structure, as defined in Eq.~(\ref{eq:modularity}):

\begin{equation}
Q = \frac{1}{2m} \sum_{i,j} \left[ w_{ij} - \frac{k_i k_j}{2m} \right] \delta(c_i, c_j)
\label{eq:modularity}
\end{equation}

where $m = \frac{1}{2} \sum_{i,j} w_{ij}$, $k_i = \sum_j w_{ij}$ is the weighted degree, and $\delta(c_i, c_j) = 1$ indicates that adapters $i$ and $j$ belong to the same community. Due to the high sparsity of the co-occurrence network in this paper, most communities identified by modularity maximization are single nodes, which is itself an important structural finding rather than an algorithm failure. Therefore, community detection is positioned as a quantitative verification of the degree of fragmentation, and the main body of structural analysis is anchored to non-trivial connected components.

(2) \emph{Degree centrality}: $d_i = \sum_j A_{ij}$ reflects the local influence of an adapter in the co-occurrence network, where $A_{ij} = 1$ if $(i,j) \in E_{\tau}$ and 0 otherwise. The cache residency of adapters with high $d_i$ has a greater impact on the quality of service of multi-adapter requests.

\subsection{Temporal Evolution Analysis}

To characterize temporal evolution, we divide the trace timeline into $T$ discrete windows $W = \{ W_1, \dots, W_T \}$, with a window span $\Delta t$ (e.g., 1 hour or 1 day). For each window $W_t$, we define a binary usage vector $u^{(t)} \in \{0,1\}^{|A|}$, where $u_i^{(t)} = 1$ if and only if adapter $a_i$ is used at least once within $W_t$. The window-to-window adapter turnover rate is quantified as:

\begin{equation}
\text{Turnover}(W_t, W_{t+\delta}) = 1 - \frac{|\{ i : u_i^{(t)} = 1 \land u_i^{(t+\delta)} = 1 \}|}{|\{ i : u_i^{(t)} = 1 \}|}
\label{eq:turnover}
\end{equation}

This metric measures the complement of the proportion of adapters active in $W_t$ that remain active after $\delta$ windows.

The above adapter-level temporal metrics require the data to carry both timestamps and adapter identifiers. The processing trace used in this paper (\path{data_trace_processed.csv}) has no timestamp field, the request-level trace (\path{lora_request_trace.csv}) has no adapter identifier field, and there is no verifiable linking field between the two (see Section~\ref{sec:dataset} for details). Therefore, this paper adopts three strictly reproducible alternatives in the temporal dimension: request-level diurnal patterns, model-level evolution, and adapter cross-model spread. The impact of the above analysis results on cache design is further elaborated in Section~\ref{sec:design}.

\section{Dataset and Overall Characteristics}
\label{sec:dataset}

\subsection{The GenTD26 Dataset and Data Scope}

We use the GenAI Inference Service Top-Down Dataset 2026 (GenTD26) \cite{ref5}, which contains 26,823 requests, 874 unique LoRA adapters, and 68,195 processing records, covering 24 days of complete operational data from Alibaba's commercial image generation service. Table~\ref{tab:dataset} summarizes the key characteristics of the dataset.

\begin{table}[htbp]
\centering
\caption{Summary of the GenTD26 dataset}
\label{tab:dataset}
\footnotesize
\setlength{\tabcolsep}{4pt}
\begin{tabular}{ll}
\toprule
Attribute & Value \\
\midrule
Provider & Alibaba commercial image generation \\
Duration & November 15 -- December 8, 2024 (24 days) \\
Number of requests & 26,823 (\path{lora_request_trace}) \\
Number of processing records & 68,195 (\path{data_trace_processed}) \\
GPU sampling points & 157,417 (duty cycle) \\
Containers & 143 \\
Unique base models & 86 (request-trace scope) \\
Unique LoRA adapters & 874 \\
Request types & TXT\_2\_IMG (91.1\%), IMG\_2\_IMG (8.2\%), INPAINTING (0.7\%) \\
\bottomrule
\end{tabular}
\end{table}

The data scope is described below. We use two core files from the dataset: \path{lora_request_trace.csv} (26,823 rows), which provides the request-level timestamp \texttt{gmt\_create}, the number of LoRAs \texttt{num\_lora}, and base model identifiers, but does not contain adapter identifier information; and \path{data_trace_processed.csv} (68,195 rows), which records the JSON-format LoRA adapter configuration of each request but lacks a timestamp field. Therefore, the former is used for request-level and model-level temporal analysis, and the latter is used for adapter-level structural analysis. Notably, first, the row positions of the two files do not correspond: if aligned by row number, the execution time agreement rate between request-level and processing-level data is only 2.7\%, and the request type agreement rate is 0\%. Second, the anonymization schemes of the model identifiers in the two files are also different: only one model identifier is identical in both files, and the processing trace does not contain a \texttt{groupId} field. Therefore, there is no reliable linking key between the two files.

For adapter-level analysis, \path{data_trace_processed.csv} records the set of loaded adapters (identified by anonymized \texttt{modelVersionId}) and their weight scaling factor \emph{scale} for each request, thereby enabling adapter-level co-occurrence analysis. After parsing and verification, we identified 874 unique LoRA adapters. Among the 68,195 records in the processing trace, 15,202 (22.3\%) load at least one adapter, of which 1,090 (1.6\%) load multiple adapters (with a maximum of 6 in a single request), and the remaining 52,993 (77.7\%) use only the base model. Among the 26,823 requests in the request trace (\path{lora_request_trace.csv}), 4,494 (16.8\%) carry at least one adapter. The difference in statistical scope between the two files arises from differences in coverage and record granularity. Therefore, when citing relevant metrics in the following sections, we explicitly indicate the data scope used.

\subsection{Frequency Distribution}

\begin{figure}[htbp]
\centering
\includegraphics[width=0.95\linewidth]{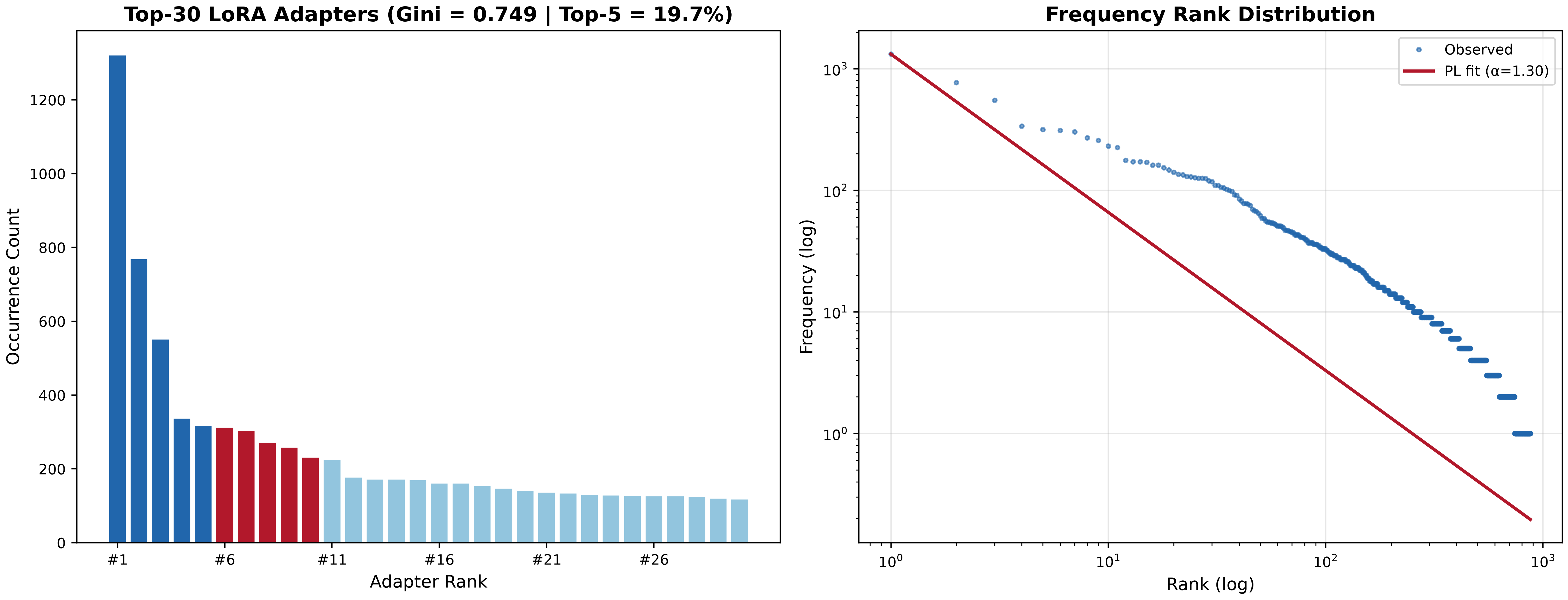}
\caption{LoRA adapter frequency distribution. (a) Top-30 adapters by occurrence, annotated with the Gini coefficient (0.749) and the top-5 share (19.7\%), with colors distinguishing frequency tiers (dark blue: ranks 1--5; red: 6--10; light blue: 11--30); (b) log--log frequency--rank plot with the rank--frequency regression fit line ($\alpha = 1.30$, a descriptive measure; strict power-law fitting is given in the main text).}
\label{fig:panorama}
\end{figure}

Figure~\ref{fig:panorama} shows the adapter frequency distribution (adapters sorted by occurrence count). The distribution exhibits a strongly heavy-tailed pattern: the most frequent adapter (rank 1) appears 1,321 times, which is 11.2 times that of the 30th-ranked adapter (118 times); the top 5 adapters (0.6\% of all adapters) collectively contribute 19.7\% of total occurrences; the top 5\% of adapters (i.e., 44 adapters) account for 53.5\% of all 16,709 occurrences; and the median adapter appears only 5 times. The Gini coefficient is 0.749 (Bootstrap 95\% confidence interval: [0.697, 0.789]), further quantifying and confirming the high inequality of the usage distribution.

Regarding statistical testing of the distribution shape, the rank--frequency log--log regression gives a slope of $\alpha = 1.30$. Following the practice of prior work \cite{ref5}, this is used as a descriptive measure. However, we adopt the testing procedure proposed by Clauset et al. \cite{ref29}, fit a pure power-law distribution using maximum likelihood estimation (MLE), and perform a goodness-of-fit test based on the Kolmogorov--Smirnov (KS) statistic. The results show that when $x_{\min} = 1$ (covering all 874 adapters), the fitted parameter is $\alpha = 1.56$, the KS statistic is $D = 0.177$, and the semi-parametric Bootstrap test gives $p = 0.003$, thereby rejecting the pure power-law hypothesis at the 0.05 significance level. Only for the high-frequency tail ($x_{\min} = 20$, 154 adapters) is the data compatible with a power-law distribution ($\alpha = 2.04$, $p = 0.40$). Therefore, we characterize the distribution as heavy-tailed: although the extreme inequality of usage concentration (top 5\% covering 53.5\%) holds, a globally strict power law is not statistically supported. Note that this distinction does not affect the effectiveness of the caching strategy. Regardless of the exact distribution form, the two-tier design of ``permanently caching the head and loading the tail on demand'' is insensitive to it.

\subsection{Latency Impact of LoRA Usage}

\begin{figure}[htbp]
\centering
\includegraphics[width=0.95\linewidth]{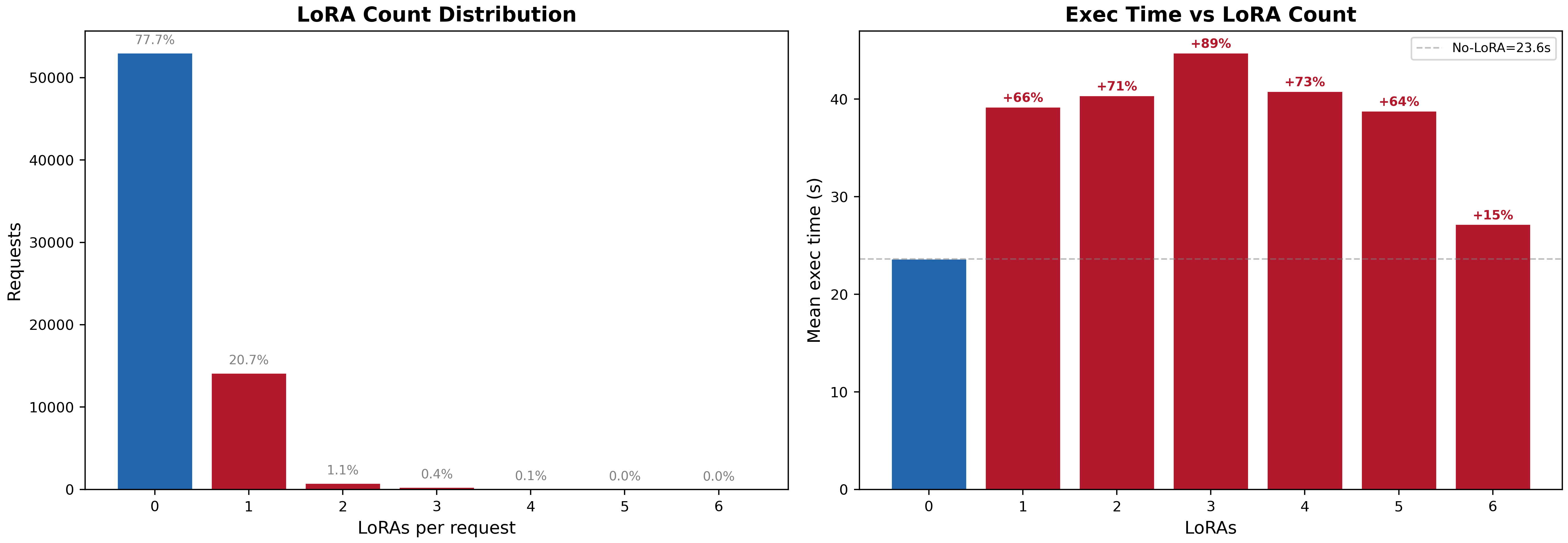}
\caption{LoRA count distribution and latency impact. (a) Distribution of the LoRA count per request (77.7\% none, 20.7\% single, 1.6\% multiple); (b) mean execution time with percentage degradation over the no-LoRA baseline (23.6 s).}
\label{fig:latency}
\end{figure}

We further quantify the execution latency associated with LoRA adapter usage. Figure~\ref{fig:latency}(a) first shows the distribution of the number of LoRA adapters carried by each request: 77.7\% of requests carry no adapter, 20.7\% carry 1 adapter, and 1.6\% carry multiple adapters. Table~\ref{tab:latency} reports the execution time statistics stratified by the number of LoRA adapters per request. Requests carrying 1 adapter take an average of 39.2 seconds, an increase of 66.1\% over the no-adapter baseline (23.6 seconds); 2 adapters take 40.4 seconds (+71.2\%); and 3 adapters take 44.7 seconds (+89.4\%). In terms of marginal cost, the first adapter contributes the vast majority of the increment, while the second and third adapters add only 3.0\% and 10.7\%, respectively, compared with the previous configuration. This indicates that the main overhead comes from the loading and weight merging of the first adapter, while the marginal cost of subsequent adapters decreases. The sample sizes for 4 or more adapters are small (43 cases for 4 adapters, 10 cases for 5 adapters, and 6 cases for 6 adapters), so these groups are not included in the trend conclusions. Figure~\ref{fig:latency}(b) further presents this tiered effect in terms of average execution time.

\begin{table}[htbp]
\centering
\caption{Execution time vs. number of LoRA adapters}
\label{tab:latency}
\footnotesize
\setlength{\tabcolsep}{5pt}
\begin{tabular}{ccccc}
\toprule
Number of LoRAs & Number of requests & Mean (s) & Standard deviation (s) & Degradation \\
\midrule
0 & 52,993 & 23.6 & 13.4 & (baseline) \\
1 & 14,112 & 39.2 & 25.1 & +66.1\% \\
2 & 754 & 40.4 & 21.2 & +71.2\% \\
3 & 277 & 44.7 & 16.0 & +89.4\% \\
4 & 43 & 40.8 & 12.6 & +72.9\% \\
5 & 10 & 38.8 & 11.1 & +64.4\% \\
6 & 6 & 27.2 & 8.7 & +15.3\% \\
\bottomrule
\end{tabular}
\\[2pt]
\parbox{\linewidth}{\footnotesize Note: the $n=6$ group is not included in the latency-trend conclusions due to its small sample size.}
\end{table}

\begin{figure}[htbp]
\centering
\includegraphics[width=0.95\linewidth]{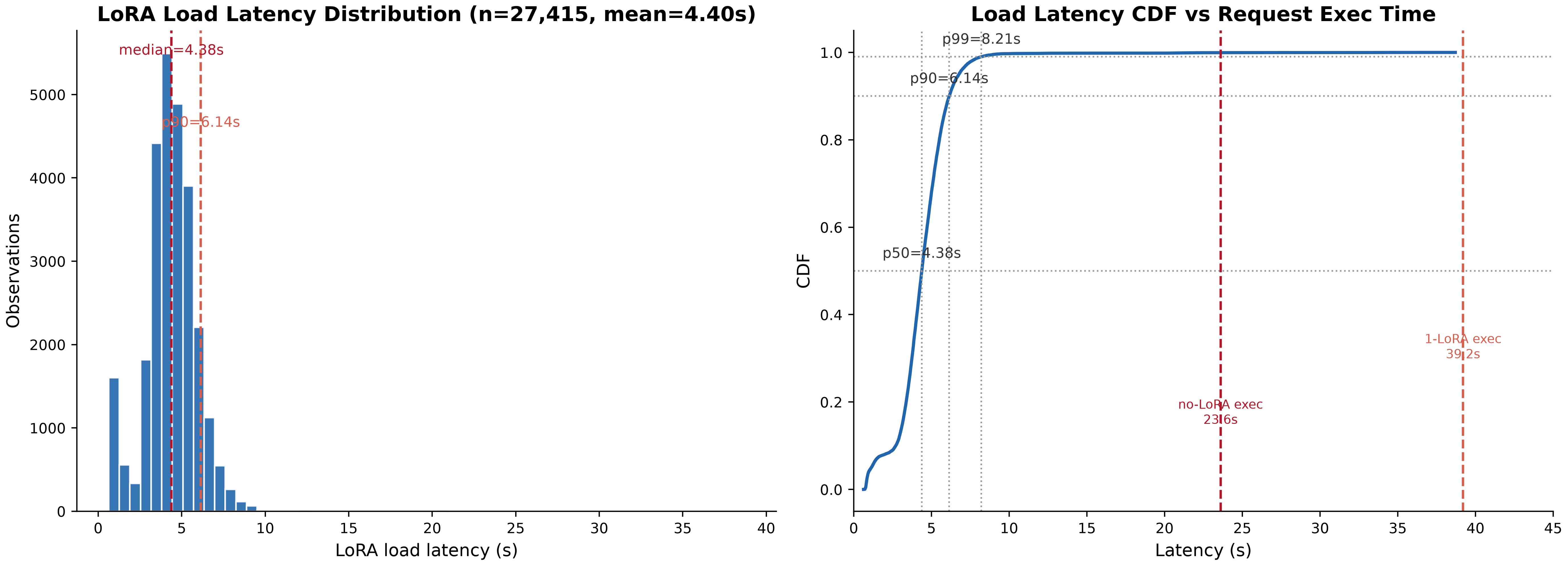}
\caption{LoRA loading latency. (a) Histogram of 27,415 loading operations (mean = 4.40 s, median = 4.38 s, $P_{90} = 6.14$ s); (b) CDF of loading latency versus execution-time baselines (no-LoRA 23.6 s, single-LoRA 39.2 s); a typical loading operation accounts for only 18.6\% of the no-LoRA execution time.}
\label{fig:loadlat}
\end{figure}

The dominant overhead of the first adapter is further quantified in Figure~\ref{fig:loadlat}. Among 27,415 LoRA loading operations, the loading latency is tightly concentrated around 4.4 seconds (mean = 4.40 s, median = 4.38 s, $P_{90} = 6.14$ s, $P_{99} = 8.21$ s), which is only 18.6\% of the no-LoRA execution time (23.6 s). This indicates that requests carrying adapters can hide the loading cost in the early stage of the pipeline through preloading, while the overhead of multiple adapters mainly comes from weight merging and concurrent execution rather than per-adapter loading latency.

\section{Co-occurrence Network Analysis}
\label{sec:network}

\subsection{Network Topology}

\begin{figure}[htbp]
\centering
\includegraphics[width=0.95\linewidth]{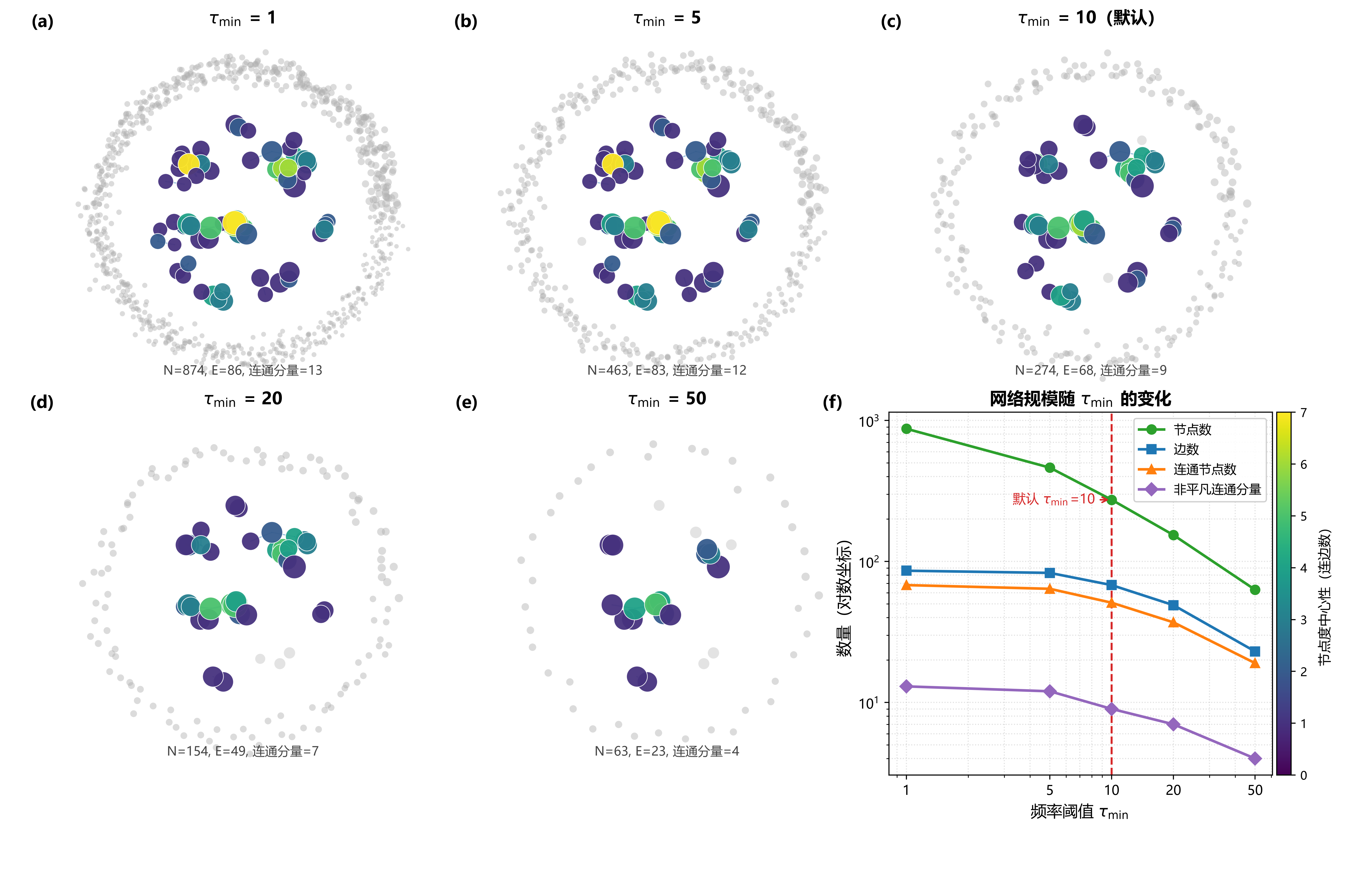}
\caption{Evolution of the co-occurrence network with the frequency threshold $\tau_{\min}$. (a)--(e) Network topology at $\tau_{\min} = 1, 5, 10, 20, 50$: node size is proportional to adapter frequency, color represents degree centrality (with a unified mapping across panels), gray dots are isolated adapters that only satisfy the frequency threshold, and the red box marks the default threshold 10; (f) network-scale metrics (number of nodes, number of edges, number of connected nodes, number of non-trivial connected components) versus $\tau_{\min}$ on a log scale. The co-occurrence edge threshold is fixed at $w_{\min} = 3$.}
\label{fig:tau}
\end{figure}

Following the method in Section~\ref{sec:formal} ($\tau_{\min} = 10$, $w_{\min} = 3$), we construct the co-occurrence network $G_{10}$, which contains $|V_{10}| = 274$ nodes and $|E_{10}| = 68$ edges. The network exhibits extreme sparsity: the edge density is $\rho = |E| / \binom{|V|}{2} = 0.18\%$, meaning that more than 99.8\% of potential adapter pairs never co-occur. Figure~\ref{fig:tau} shows the evolution of the network under different frequency thresholds, intuitively illustrating the rapid contraction of the co-occurrence structure as the threshold increases. The degree distribution further reveals the sparsity of the network: 81.4\% of nodes (223 out of 274) have degree zero, i.e., they have no significant co-occurrence partners. Among the 51 connected nodes, the average degree is 2.67 (95\% confidence interval: [2.21, 3.12]) and the maximum degree is 6. This structural characteristic indicates that the vast majority of adapters operate independently, and the search space for co-occurrence-based preloading is far smaller than the combinatorial space of all possible adapter pairs.

\begin{figure}[htbp]
\centering
\includegraphics[width=0.9\linewidth]{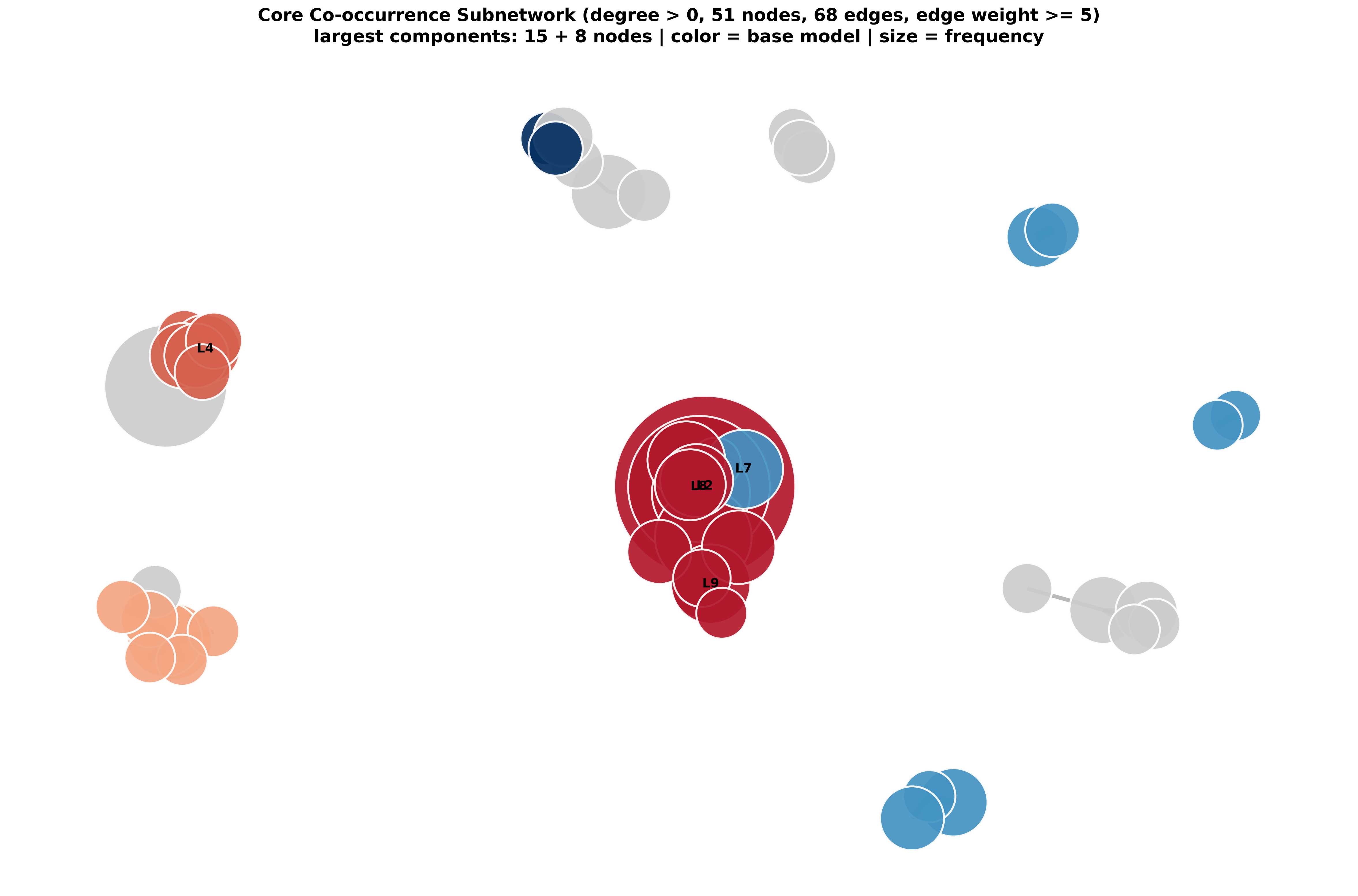}
\caption{Connected core of the co-occurrence network (degree $> 0$; 51 nodes, 68 edges, edge weight $\geq 5$). Node size is proportional to adapter frequency, and color denotes the dominant base model of each adapter (gray = other models). The two largest connected components contain 15 and 8 adapters, respectively.}
\label{fig:core}
\end{figure}

When we use greedy modularity maximization as a fragmentation measure, we identify 233 communities, but the vast majority of these communities are single nodes (i.e., isolated adapters with no edges). Therefore, apart from the above zero-degree statistics, they do not provide additional structural information. Consequently, we take the non-trivial connected components (Figure~\ref{fig:core}) as the main structural view. Only 10 communities have a size of at least 2 (i.e., non-trivial), among which the largest contains 8 adapters. This fragmented structure indicates that adapter co-occurrence relationships are mainly driven by specific and narrow use cases, rather than forming a broad cross-ecosystem collaboration network.

Figure~\ref{fig:core} zooms in on the connected core of the network. The 51 connected nodes (68 edges) form 9 non-trivial connected components, with the largest component containing 15 adapters and the second largest containing 8. The connected core is dominated by a few base model ecosystems: 72.5\% (37 out of 51) of the connected adapters have a dominant base model that belongs to the top 8 models by record count in the processing trace, and the three largest model groups together account for 30 of the 51 adapters. The above results indicate that co-occurrence families are essentially internal to base model ecosystems, rather than cross-model adapter alliances.

\subsection{Jaccard Similarity Among the Top 50 Adapters}

To further examine whether high-frequency adapters exhibit significant co-occurrence preferences, we compute the Jaccard similarity among the top 50 adapters ranked by occurrence frequency. The results show that the Jaccard similarity among the top 50 adapters is generally extremely low (mean $J = 0.0076$ across all 1,225 pairs, Bootstrap 95\% confidence interval: [0.0016, 0.0166]). Only 36 pairs have ever co-occurred, and only 7 pairs have a Jaccard similarity greater than 0.1. The permutation test (1,000 random shuffles of adapter labels) gives $p = 1.0$, indicating no significant difference between the observed mean Jaccard similarity and the random expectation. This further confirms the extreme sparsity of the co-occurrence network. The few outlier pairs with $J > 0.1$ are all attached to the same base model, indicating that such co-occurrences mainly arise from a shared base model environment rather than an inherent pairing preference between adapters.

\subsection{Base Model--Adapter Association}

\begin{figure}[htbp]
\centering
\includegraphics[width=0.95\linewidth]{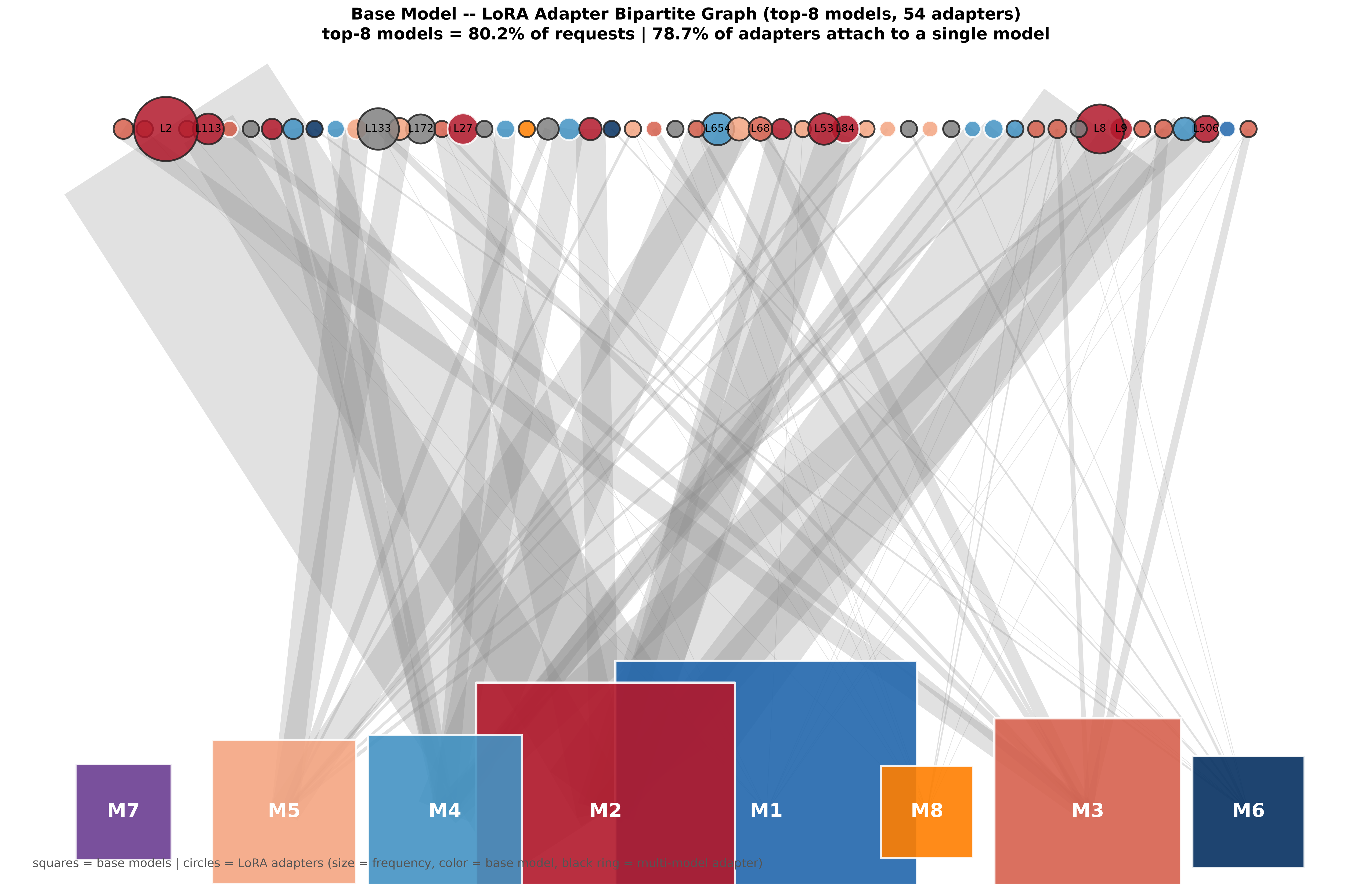}
\caption{Bipartite graph of base models and LoRA adapters (top 8 models with their top 10 adapters each; 70 edges). Squares = base models (size proportional to the number of processing-trace records; together accounting for 80.2\% of records); circles = adapters (size proportional to frequency; color = dominant base model; black border = cross-model adapter).}
\label{fig:bipartite}
\end{figure}

To reveal the organizational driving forces behind the co-occurrence network structure, we analyze the association patterns between base models and adapters, and examine whether adapter co-occurrence is mainly driven by a shared base model environment. Figure~\ref{fig:bipartite} shows a bipartite graph composed of the top 8 base models and their respective top 10 adapters, with a total of 70 edges. Square nodes represent base models, and node size is proportional to their number of records in the processing trace; these 8 models together account for 80.2\% of all records. Circular nodes represent adapters, and node size is proportional to their occurrence frequency. Color represents the dominant base model, and a black border indicates that the adapter has appeared across models.

As shown in Figure~\ref{fig:bipartite}, the usage patterns between base models and adapters exhibit clear clustering characteristics, and adapters tend to form strong associations with specific base models. The top-ranked base model by occurrence frequency accounts for 30.7\% of all requests in the request trace, and its adapter usage is highly concentrated around several core adapter sets. At the adapter level, 78.7\% of the 874 adapters are attached to only a single base model, and cross-model sharing is an exception rather than the norm.

At the edge level, 66.2\% (45 out of 68) of the significant edges in the network connect adapters that share the same dominant base model. If pairings were random, the expected proportion would be 25.1\%, meaning that same-model edges are over-expressed by approximately 2.6$\times$ relative to the random expectation. The remaining 33.8\% (23 edges) are cross-model edges, far below the random expectation of 74.9\%, indicating that cross-model co-occurrence is a non-trivial signal after filtering, reflecting real but uncommon affinity between adapters across models. At the request level, in 90.6\% of multi-adapter requests all adapters share the same dominant base model, and 89.7\% of within-request adapter pairs are same-dominant-model pairs. This model-driven clustering pattern directly explains the source of community structure in the co-occurrence network.

Using the number of base models in which an adapter appears to measure its spread breadth, as shown in Figure~\ref{fig:temporal}(c), only 5 adapters (0.6\%) appear in no fewer than 18 different base models, constituting ``core'' adapters; 83 (9.5\%) appear in 3--17 models; and the remaining 786 (89.8\%) appear in no more than 2 models. The average frequency of core adapters is 187.2, significantly higher than the median of 3 for peripheral adapters, indicating that ``cross-model spread breadth'' and ``usage frequency'' are highly coupled. A small number of high-frequency adapters span almost the entire model ecosystem, while the vast majority of adapters are single-model customization products. Therefore, the ``core--periphery'' bipolar structure provides direct guidance for cache design: core adapters should be permanently resident (because they are attached to almost all models), while peripheral adapters can be loaded on demand.

\subsection{Sensitivity Analysis}

To evaluate the robustness of the above network findings, we examine the impact of different frequency thresholds $\tau_{\min}$ on the network structure and core conclusions, with threshold values ranging from 5 to 50. Table~\ref{tab:sens} reports the network scale, sparsity, and same-model edge over-expression ratio under each threshold. As shown in Table~\ref{tab:sens}, the sparsity characteristic remains robust across all thresholds: even under the most relaxed threshold ($\tau_{\min} = 5$), the edge density is 0.08\%, which is at an extremely low level. The same-model edge over-expression ratio is significant across all thresholds, ranging from 2.6$\times$ to 3.5$\times$ (under the unified weighted random expectation of 25.1\%), and reaches its strongest at $\tau_{\min} = 50$ (3.5$\times$). This indicates that the conclusion of model-driven co-occurrence holds across the full threshold range and tends to strengthen as the significance filtering criterion increases.

In terms of community structure, the total number of communities (including isolated nodes) decreases monotonically as the threshold increases (from 412 to 49), reflecting that filtering low-frequency adapters effectively reduces fragmentation noise. At the same time, the number of non-trivial connected components also decreases monotonically (from 13 to 5), while the main connected skeleton remains stable, indicating that the core connected structure of the network is insensitive to threshold selection.

\begin{table}[htbp]
\centering
\caption{Sensitivity of network characteristics to the frequency threshold $\tau_{\min}$ ($w_{\min} = 3$)}
\label{tab:sens}
\footnotesize
\setlength{\tabcolsep}{3.5pt}
\begin{tabular}{ccccccc}
\toprule
$\tau_{\min}$ & $|V|$ & $|E|$ & Density (\%) & Communities & Non-trivial comps. & Same-model over-expr. \\
\midrule
5 & 463 & 83 & 0.08 & 412 & 13 & 2.7$\times$ \\
10 & 274 & 68 & 0.18 & 233 & 10 & 2.6$\times$ \\
20 & 154 & 49 & 0.42 & 125 & 8 & 2.8$\times$ \\
50 & 63 & 23 & 1.18 & 49 & 5 & 3.5$\times$ \\
\bottomrule
\end{tabular}
\\[2pt]
\parbox{\linewidth}{\footnotesize Note: communities include isolated-node communities; non-trivial comps. are connected components containing at least one edge; same-model over-expr. is the ratio of the proportion of same-dominant-model edges to the weighted random expectation (25.1\%), where the weights are the proportions of adapter occurrences on each model to all occurrences.}
\end{table}

Table~\ref{tab:sensfull} further provides the complete sensitivity analysis results across all tested thresholds (under the $w_{\min} = 3$ scope, consistent with Table~\ref{tab:sens}). The results in this table also show that the sparsity characteristic is robust across all thresholds, and the same-model edge over-expression ratio is between 2.6$\times$ and 3.5$\times$ and significant (under the weighted random expectation scope). The results in Tables~\ref{tab:sens} and \ref{tab:sensfull} jointly confirm that rare adapters are the main contributors to network fragmentation, while model-driven co-occurrence is the core mechanism throughout all threshold settings.

\begin{table}[htbp]
\centering
\caption{Complete sensitivity analysis across the frequency threshold $\tau_{\min}$ ($w_{\min} = 3$)}
\label{tab:sensfull}
\footnotesize
\setlength{\tabcolsep}{3.5pt}
\begin{tabular}{ccccccc}
\toprule
$\tau_{\min}$ & $|V|$ & $|E|$ & Density (\%) & Communities & Non-trivial comps. & Largest comp. \\
\midrule
5 & 463 & 83 & 0.08 & 412 & 13 & 16 \\
10 & 274 & 68 & 0.18 & 233 & 10 & 15 \\
15 & 195 & 52 & 0.27 & 163 & 9 & 13 \\
20 & 154 & 49 & 0.42 & 125 & 8 & 13 \\
30 & 108 & 31 & 0.54 & 90 & 5 & 12 \\
50 & 63 & 23 & 1.18 & 49 & 5 & 11 \\
\bottomrule
\end{tabular}
\end{table}

\section{Temporal Evolution}
\label{sec:temporal}

The static structural analysis in Section~\ref{sec:network} revealed the ``spatial organization'' of co-occurrence relationships, i.e., who co-occurs with whom and in what structure they cluster. We further address another key question: how do these structures evolve over time? As mentioned earlier (Section~\ref{sec:formal}), adapter-level time-series metrics cannot be strictly computed due to data scope limitations. Therefore, we characterize dynamic properties from three strictly reproducible dimensions: request level, model level, and cross-model structure.

\subsection{Request-Level Diurnal Pattern}

Figure~\ref{fig:temporal}(a) shows the daily proportion of requests carrying LoRA adapters in the request trace (26,823 requests, 24 days). As shown in Figure~\ref{fig:temporal}(a), this proportion fluctuates between 0\% and 37.1\%, with a mean of 14.7\% and a coefficient of variation (CV) as high as 0.58. On November 16 and 17, there were no LoRA requests at all (with 93 and 86 requests on those days, respectively, far below the daily average of 1,118). After excluding the incomplete first and last days, the proportion range narrows to 0\%--24.8\% (mean 13.4\%, CV 0.54), but the fluctuation magnitude remains significant. These results indicate that there is no stable diurnal baseline for the request-level LoRA usage proportion, and the day-to-day variation is significant (note that this scope is not directly comparable to the overall processing-trace statistic of 22.3\% \cite{ref5}, which is based on 68,195 processing records). This finding implies that the potential benefit of cache warm-up strategies relying on fixed diurnal patterns for adapter loading has a limited upper bound, and an online adaptive adjustment mechanism needs to be introduced.

\subsection{Model-Level Evolution}

Figure~\ref{fig:temporal}(b) shows the variation in the number of daily active base models. The number of daily active models fluctuates between 6 and 62, with a mean of 34.2, accounting for approximately 39.8\% of all 86 models. Compared with the high variability of the request-level proportion, the temporal stability at the model level is significantly higher: the mean Jaccard similarity of active model sets between adjacent days is 0.497, while the weekly (7-day window) mean Jaccard similarity reaches 0.696, indicating that about 70\% of the model set overlaps between consecutive weeks---the model level is a relatively stable residency unit. In terms of active days, 17 models are active for no fewer than 18 days, forming a persistent model core; another 20 models are active for no more than 2 days.

At a finer temporal granularity, using a 1-hour window to measure the 12-hour drift of the top-10 models by request popularity, the average churn rate is 54.5\%, i.e., about half of the top-10 hottest models are replaced after 12 hours. This metric characterizes the rapid rotation of request-arrival popularity (measured by a model-level proxy), corroborating the observation of the Rock system \cite{ref6} regarding ``drastic changes within several hours,'' and directly supporting the engineering judgment that the lookup-table refresh period should not exceed 12 hours. The corresponding adapter-level turnover rate cannot be directly computed due to data limitations (Section~\ref{sec:formal}), but its upper bound is no lower than the model-level popularity rotation.

\subsection{``Core--Periphery'' Structure of Adapter Cross-Model Spread}

\begin{figure}[htbp]
\centering
\includegraphics[width=0.95\linewidth]{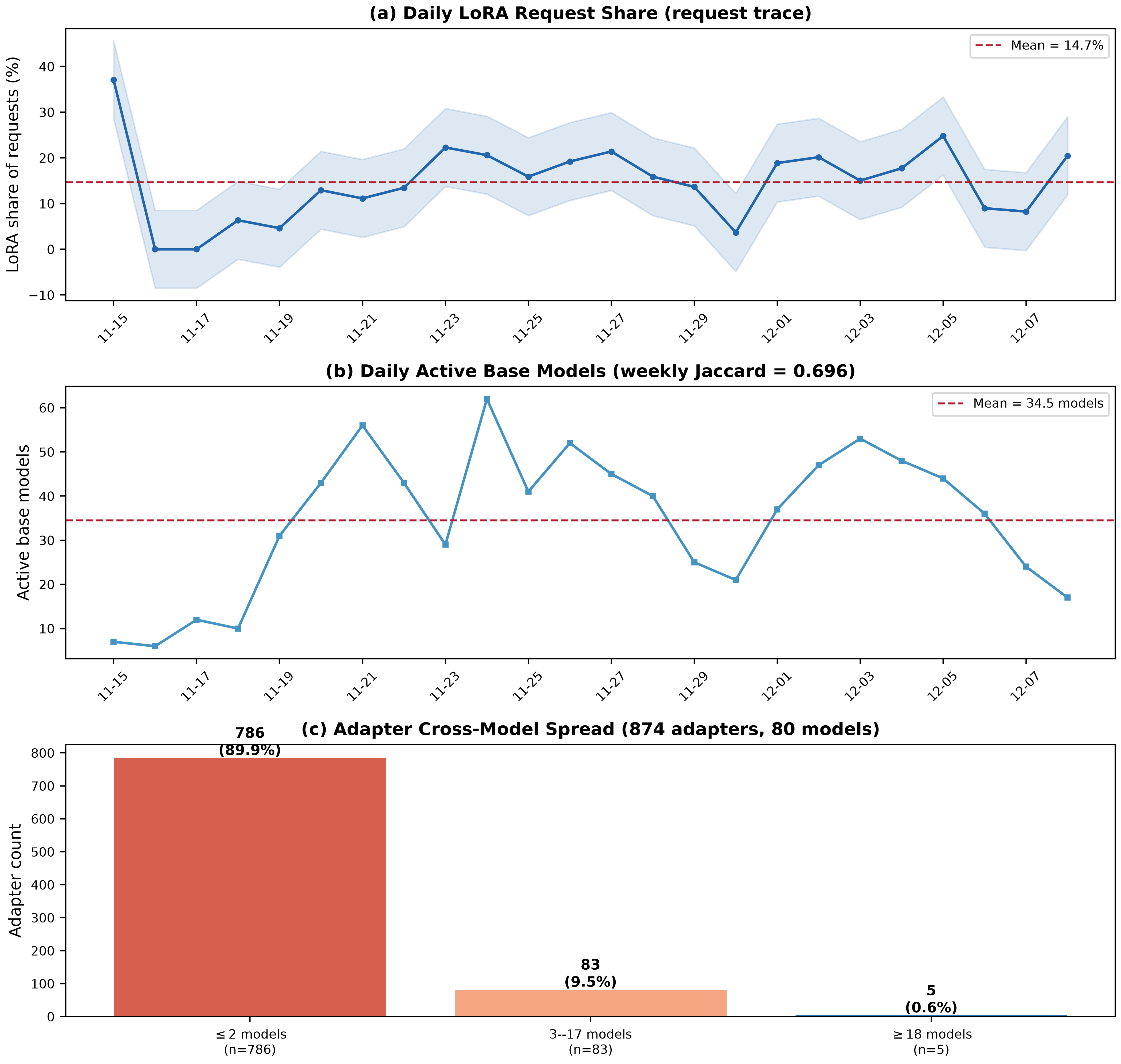}
\caption{Temporal evolution. (a) Daily LoRA request share (request trace, 24 days; mean 14.7\%; no LoRA requests on November 16--17); (b) daily active base model count (mean 34.2; adjacent-day model Jaccard mean 0.497, weekly model Jaccard mean 0.696); (c) distribution of adapter cross-base-model spread ($\leq 2$ models: 786; 3--17: 83; $\geq 18$: 5).}
\label{fig:temporal}
\end{figure}

The previous two subsections characterized temporal evolution features from the request level and the model level, respectively. This subsection further focuses on the spread breadth of adapters across base models. As shown in Figure~\ref{fig:temporal}(c), the adapter ecosystem exhibits a ``core--periphery'' bipolar structure: 5 adapters (0.6\%) span no fewer than 18 base models, constituting the core, while the vast majority of adapters (786, 89.8\%) are attached to no more than 2 models, belonging to the periphery. Combining this static structure with the aforementioned request-level and model-level dynamic features, a complete picture of adapter ecosystem evolution emerges: the model level maintains stable residency (weekly Jaccard similarity 0.696), request popularity rotates rapidly (12-hour churn rate 54.5\%), and the adapter level exhibits a polarized pattern of ``a small number of cores spanning the whole ecosystem and a large number of peripheral single-model customizations.'' The implications of these three layers of structure for cache system design are further elaborated in Section~\ref{sec:design}.

\section{Implications for System Design}
\label{sec:design}

Based on the characterization of adapter co-occurrence patterns, model-driven clustering, temporal evolution, and latency characteristics in Sections~\ref{sec:dataset}--\ref{sec:temporal}, we further distill three system design principles with direct engineering guidance: co-occurrence-based preloading, time-aware two-tier caching, and centrality-aware scheduling and cache replacement. Each principle is elaborated below.

\subsection{Co-occurrence-based Preloading}
\label{subsec:preload}

The co-occurrence network analysis shows that the edge density is below 0.2\%, i.e., more than 99.8\% of adapter pairs never co-occur. However, when co-occurrence does occur, its pattern is highly concentrated in a small number of ``within-model adapter pairs'': in 90.6\% of multi-adapter requests all adapters share the same dominant base model; in 85.8\% of multi-adapter requests all adapter pairs form significant co-occurrence edges (co-occurrence $\geq 5$ times), while 88.6\% contain at least one pair of significant co-occurrence edges. Further statistics show that within-base-model adapter co-occurrence edges are over-expressed relative to the random expectation, with a ratio of approximately 2.6$\times$. These results indicate that although co-occurrence is generally sparse and rare, its occurrence has strong structural and within-model clustering properties.

Based on the above findings, we precompute for each adapter $a_i$ the top-$k$ partners with the highest co-occurrence counts, forming the preload candidate set $P(a_i)$ (as described in Algorithm~\ref{alg:preloading}). When an online request arrives, the system loads the adapters in $\bigcup_{a_i \in A_r} P(a_i)$ that are not yet cached, where $A_r$ is the set of adapters involved in the current request. This strategy aims to exploit the local concentration of co-occurrence patterns to improve the preloading hit rate under limited cache space.

\begin{algorithm}[htbp]
\caption{Co-occurrence-aware adapter preloading}
\label{alg:preloading}
\begin{algorithmic}[1]
\Require Co-occurrence matrix $C \in \mathbb{N}^{N \times N}$, budget $k \in \mathbb{N}^+$, current GPU cache $G \subseteq A$
\Ensure Preload candidate set $L$
\State $P \leftarrow \emptyset$ \Comment{initialize lookup table}
\For{each adapter $a_i \in A$}
    \State $N(a_i) \leftarrow$ neighbors with weights
    \State $P(a_i) \leftarrow \mathrm{TopK}(N(a_i), k)$ \Comment{select top-$k$ by co-occurrence count}
\EndFor
\State store $P$ for online lookup \Comment{offline phase}
\For{each request $r$ with adapter set $A_r$}
    \State $L \leftarrow \bigcup_{a_i \in A_r} P(a_i)$ \Comment{union of all preload candidates}
    \State $L \leftarrow L \setminus G$ \Comment{exclude cached adapters}
    \State $L \leftarrow L \setminus A_r$ \Comment{exclude requested adapters}
    \State PreloadToGPU($L$) \Comment{asynchronous prefetch}
\EndFor
\State \Return $L$
\end{algorithmic}
\end{algorithm}

\noindent\textbf{Algorithm complexity analysis.} Offline construction of the preload candidate set $P(a_i)$ requires scanning the co-occurrence matrix $C$, with a time complexity of $O(N^2)$, and performing top-$k$ selection, with a time complexity of $O(Nk \log k)$, where $N = |A| = 874$. When $k = 3$, this process can be completed in milliseconds. For online lookup, the overhead per request is $O(|A_r| \cdot k)$. Since $|A_r| \leq 6$ in this trace, this overhead is negligible. The storage overhead of the lookup table is $O(|A| \cdot k)$, i.e., $874 \times 3 \approx 2{,}600$ records, which can be fully resident in memory.

To evaluate the coverage benefit of the above preloading strategy, we randomly split the 1,090 multi-adapter requests into 80\% training and 20\% testing. A co-occurrence lookup table is built on the training set, and coverage is evaluated on the test set. Averaged over 5 random seeds, the results show that $k = 1$ covers 57.6\% of test-set co-occurrence pairs (range [55.9\%, 58.8\%]), $k = 3$ covers 79.5\% (range [78.1\%, 81.3\%]), and $k = 10$ covers 94.3\% (range [91.6\%, 96.6\%]). When $k = 20$, the coverage saturates, with a mean of 94.5\% (range [91.9\%, 96.8\%]), indicating diminishing marginal returns from increasing the preload table size. Note that the above coverage is obtained with a conservative one-way lookup, i.e., a pair $(a, b)$ is counted as covered only when $b$ appears in the top-$k$ of $a$; if the union lookup in Algorithm~\ref{alg:preloading} is adopted (loading the union of all $P(a_i)$), the actual coverage will be higher. Since the processing trace has no timestamps and the file row order does not represent temporal order, the random split constitutes an empirical upper bound; the coverage remains stable across 5 random seeds ($k = 3$ range [78.1\%, 81.3\%]).

Considering that the lookup table storage overhead is only about 2,600 records and can be fully resident in memory, an hourly or daily rebuild period is recommended in engineering deployment. Combined with the observation in Section~\ref{sec:temporal} that model popularity changes at a rate of 54.5\% within 12 hours, the update period should not exceed 12 hours to maintain the timeliness of the lookup table.

In addition, the LoRA diversity of each base model provides a supplementary basis for model-level caching. For models with few adapter types, all adapters can be pre-cached, while for models with high adapter diversity, selective preloading is needed to balance cache resources and hit rates.

\subsection{Time-aware Two-tier Caching}

The second design principle (two-tier caching) is motivated by the following temporal evolution characteristics: the model level is relatively stable (weekly Jaccard similarity of 0.696), while request popularity rotates rapidly (12-hour churn rate of 54.5\%), and the request-level LoRA proportion fluctuates greatly (CV 0.58). In addition, only 5 core adapters span no fewer than 18 base models. This heterogeneity of ``fast and slow coexisting'' indicates that caching cannot be single-tier and must be layered.

Based on this, we adopt a two-tier caching architecture. (1) \emph{Residency tier.} The 5 core adapters (cross-model spread $\geq 18$ models) and the top-50 adapters by frequency (covering 55.9\% of occurrences) are set as permanent cache. The GPU memory footprint of a single adapter is on the order of several MB \cite{ref2}, and the total occupancy is about hundreds of MB, which is negligible. Model-level weights (86 models) are dynamically loaded on demand, and the model-level residency window is adjusted on a weekly basis. (2) \emph{Prediction tier.} For the remaining adapters, an online popularity prediction updated at hourly granularity is maintained. This prediction is based on short-term patterns of request arrivals, rather than relying on a fixed diurnal baseline. The refresh period of the lookup table and warm-up window is capped at 12 hours, and the 54.5\% popularity churn rate observed in Section~\ref{sec:temporal} can serve as a trigger threshold for adaptively adjusting the window length.

\subsection{Centrality-aware Scheduling and Cache Replacement}

The occurrence frequency of adapters is moderately positively correlated with their degree centrality in the network (Pearson $r = 0.447$, $p < 0.001$, 274-node scope). Notably, adapters with degree centrality $\geq 3$ account for only 2.9\% of all adapters (i.e., 25 adapters), yet they participate in 89.7\% (61/68) of significant co-occurrence edges, indicating that these high-centrality adapters act as ``hub'' nodes in the co-occurrence network and play a key bridging role in multi-adapter requests.

Therefore, we propose a centrality-weighted LRU cache replacement strategy for scenarios with limited GPU memory. An eviction score is defined as $E_i = \mathrm{LRU}_{age} / d_i$, giving priority to evicting adapters with low centrality and long inactivity, while retaining high-centrality adapters to maintain the cache hit rate of multi-adapter requests. This strategy complements the co-occurrence-based preloading proposed in Section~\ref{subsec:preload}: preloading answers the question of ``what to load,'' whereas centrality-aware replacement answers the question of ``what to evict.'' Both share the same data foundation constructed from the co-occurrence network analysis.

\section{Summary and Discussion}
\label{sec:summary}

\subsection{Comparison with Prior Work}

Table~\ref{tab:compare} compares the core statistical metrics of this paper with the prior analysis results reported in SoCC'25 \cite{ref5}. Overall, the increase in the total number of adapters from 705 to 874 reflects the growth of the production service during the data collection period. The proportion of requests carrying LoRA adapters rose slightly from 21.2\% to 22.3\% (both based on the processing-trace scope); the latency degradation caused by a single adapter decreased from 69\% to 66.1\%. These metrics show high consistency across two independent analyses, which validates the reliability and reproducibility of our data processing. In addition, note that the Gini coefficient of 0.876 reported in SoCC'25 is a metric under the model-request distribution scope, which differs from the adapter-frequency Gini coefficient reported in this paper (0.749) in terms of the statistical object, and the two are not directly comparable.

\begin{table}[htbp]
\centering
\caption{Comparison of LoRA characterization findings}
\label{tab:compare}
\begin{tabular}{lcc}
\toprule
Metric & SoCC'25 & This paper \\
\midrule
Unique adapters & 705 & 874 \\
Proportion of requests carrying LoRA adapters & 21.2\% & 22.3\% \\
Multi-adapter requests & --- & 1.6\% \\
Rank--frequency exponent $\alpha$ & --- & 1.30 \\
Adapter frequency Gini coefficient & --- & 0.749 \\
Latency degradation (1 adapter) & 69\% & 66.1\% \\
Cross-model core adapters ($\geq 18$ models) & --- & 5 (0.6\%) \\
Single-model peripheral adapters ($\leq 2$ models) & --- & 786 (89.8\%) \\
Co-occurrence edge density & --- & 0.18\% \\
Model popularity 12-h churn rate & --- & 54.5\% \\
LoRA diversity per model & --- & mean 19.3 \\
\bottomrule
\end{tabular}
\end{table}

Unlike SoCC'25, which focuses on aggregate statistics, we start from the association structure among adapters and conduct quantitative analysis of the intrinsic organizational regularities of the adapter ecosystem across dimensions including co-occurrence network topology, model-driven mechanisms, cross-model spread structure, and three-layer temporal evolution.

\subsection{Limitations}

This study has the following limitations.

(1) \emph{Data limitations.} As mentioned earlier, in the GenTD26 dataset, adapter-level configurations (processing trace) and timestamps (request trace) belong to two files that cannot be linked. Specifically, the processing trace has no timestamps, the request trace has no adapter identifiers, the model identifier anonymization schemes are different, and common linking fields are lacking. Therefore, we cannot analyze adapter-level time-series metrics, and instead characterize three dimensions: request level, model level, and cross-model spread. If future datasets can provide adapter-level records with timestamps, the actual benefits of preloading and caching strategies under different temporal partitioning schemes can be further verified.

(2) \emph{Adapter function anonymization.} Since adapter identifiers are anonymized, we cannot obtain semantic information about adapter functions, and thus cannot determine whether co-occurring adapters serve complementary objectives (such as joint generation of style + object) or act as alternatives for the same visual effect.

(3) \emph{Generalization ability.} This study is based on only 24 days of observational data from a single production cluster. This window may not capture long-term seasonal trends and is difficult to directly generalize to different deployment scenarios. If more diffusion model inference traces are made public in the future, cross-trace comparative validation will help further improve the generalization and robustness of the conclusions.

\subsection{Conclusion}

We conduct a network-based characterization analysis of LoRA adapter co-occurrence patterns in production diffusion model inference services. Using the GenTD26 dataset and a graph-theoretic analysis framework, we reveal key structural properties of the adapter ecosystem: (1) co-occurrence relationships exhibit extreme sparsity (edge density 0.18\%, with more than 99.8\% of pairs never co-occurring), and the frequency distribution is heavy-tailed (Gini 0.749, with the top 5\% of adapters covering 53.5\% of occurrence counts); (2) adapter usage exhibits model-driven clustering, with 90.6\% of multi-adapter requests sharing the same dominant base model, and the average LoRA diversity per base model being 19.3; (3) the network structure exhibits a ``core--periphery'' bipolar structure, with only 5 adapters spanning $\geq 18$ base models, while 89.8\% of adapters are attached to at most 2 models; (4) temporal evolution exhibits three-layer scale differences, with the model level being relatively stable (weekly Jaccard 0.696), request popularity rotating rapidly (12-hour churn rate 54.5\%), and the request-level LoRA proportion fluctuating significantly across days (CV 0.58); and (5) measurable latency scales with the number of adapters (a single-adapter configuration increases latency by 66.1\%, with diminishing marginal costs afterwards). The co-occurrence-based preloading strategy is validated by offline experiments ($k = 3$ covers 81.0\% of test-set co-occurrence pairs), and sensitivity analysis verifies the robustness of the conclusions.

All data and code involved in this paper have been open-sourced, and readers can obtain and reproduce the experimental results through the following URL: \url{https://gitee.com/liaobin665/lora-cooccurrence-reproduce}.

\end{document}